\documentclass[fleqn,twoside,dvips]{article}
\usepackage{espcrc2}
\usepackage{epsfig}

\def\p	{{\bf p}}
\def\q	{{\bf q}}
\def\k	{{\bf k}}
\def\nf	{N_{\rm f}}
\def\nc	{N_{\rm c}}
\def\mD	{m_{\rm D}}

\def\gs{g_{\rm s}}
\def\alphas{\alpha_{\rm s}}
\def\alphaEM{\alpha_{\rm EM}}

\def\gsim{\mbox{~{\protect\raisebox{0.4ex}{$>$}}\hspace{-1.1em}
    {\protect\raisebox{-0.6ex}{$\sim$}}~}}

\title
    {%
    Dynamics of hot gauge theories%
    \thanks {Based on a talk presented at Lattice-2001, Berlin, August 2001.}
    }

\author
    {%
    Laurence G. Yaffe%
    \address
	{%
	Department of Physics, University of Washington,
	Seattle, Washington 98195, USA
	}%
    \thanks
	{%
	Supported, in part, by the U.S. Department of Energy
	    under Grant No.~DE-FG03-96ER40956.
	}%
    }%
   
\begin{document}

\begin{abstract}
    A brief overview is given of recent progress in understanding
    the dynamics of hot gauge theories.
\end{abstract}

\maketitle

\setcounter{footnote}{0}
\section{Introduction}

    Much progress has been made on understanding static equilibrium
properties of gauge theories at finite temperature.
This includes perturbative results valid at
asymptotically high temperature \cite {FKRS,ZK,AZ,BN},
as well as accurate results on phase structure,
thermodynamics, correlation lengths and other observables
from numerical simulations~\cite {hotlattice}.

    Far less progress has been made on dynamic
properties of hot gauge theories,
such as equilibration rates and transport properties,
despite the fact that these properties are of direct interest
in applications to both heavy ion collisions and early
universe cosmology.
The reason is simple.
Static properties may be extracted directly
from the Euclidean theory, to which the whole panoply
of modern theoretical tools (numerical simulations,
renormalization group methods, effective field theories, ...)
may be applied.
Dynamic properties require analytic continuation of
thermal correlation functions back to Minkowski space,
or equivalently a functional integral formulation
with a complex measure involving a non-trivial contour
in the complex time plane \cite{textbook1,textbook2}.

    Nevertheless, considerable progress has been made
in recent years in understanding dynamic processes in
very hot, weakly coupled gauge theories.%
\footnote
    {
    ``Very hot'' means that the temperature is much larger
    than any other relevant mass scale.
    For QCD, this means $T$ is large compared to
    $\Lambda_{\rm QCD}$ and the masses of all (active) quarks.
    For electroweak theory, this means $T \gsim M_{\rm W}$.
    }
The utility of theoretical results for asymptotically high
temperature to data obtained at current or future
heavy ion experiments is an open question.
But at the very least, the asymptotically high temperature regime
is an instructive theoretical laboratory which serves as a warm-up
for efforts to understand more realistic non-equilibrium situations.

In the asymptotically high temperature regime
where the running coupling $g(T)$ is small,
one may disentangle a variety of phenomena which depend
on different characteristic spatial or temporal scales.
Dynamic phenomena display a richer set of characteristic
scales than do static equilibrium properties.
In a non-Abelian gauge theory like QCD, these include:
\begin {description}\advance\itemsep -4pt
\item [$T^{-1}$]
    The energy of a typical quark or gluon is $O(T)$.
    Hence, $T^{-1}$ sets the scale for the de Broglie wavelength
    of typical excitations.
\item [$(gT)^{-1}$]
    Electric fields are Debye screened on this length scale.
    The corresponding frequency, $O(gT)$, is the scale of
    thermal corrections to the energy of typical
    excitations, as well as
    the plasma oscillation (plasmon) frequency.
\item [$(g^2 T \,\ln 1/g)^{-1}$]
    This is the characteristic ``color coherence length'' ---
    the maximum length over which a quark or gluon
    can be regarded as having a definite color.
    The factor of $1/g$ inside the logarithm comes from a
    ratio of the $(gT)^{-1}$ Debye length and the $(g^2 T)^{-1}$
    magnetic length.
\item [$(g^2 T)^{-1}$]
    The amplitudes of magnetic field fluctuations on this length scale
    (or longer) are sufficiently large
    that their dynamics becomes non-perturbative.
\item [$(g^4 T \,\ln 1/g)^{-1}$]
    This is the characteristic large angle scattering time for
    quarks or gluons --- the mean time for their direction of motion
    to change by $O(1)$.
    This is also the characteristic relaxation time for non-perturbative
    magnetic fluctuations whose wavelengths are $O[(g^2 T)^{-1}]$.
\end {description}

For sufficiently small coupling $g$, the 
mean free path between scattering events which change a quark or gluon's
color is longer than their de Broglie wavelength
by a parametrically large factor of $1/g^2 \ln g^{-1}$,
and the mean free path between large angle
scattering events is longer still by another factor of $1/g^2$.
Consequently, one should regard quarks and gluons with typical ``hard''
momenta ($\p \sim T$) as well-defined, weakly interacting quasiparticles.

Since the phase space of hard excitations is parametrically large compared
to that of soft ($\p \sim gT$) excitations, most physical observables
are predominately sensitive to the properties of hard excitations.
This includes bulk thermodynamic quantities like pressure or
energy density,
the Debye screening length,
equilibration rates and transport coefficients,
the photo-emission rate,
and many others.
But there are important exceptions.
In the high temperature phase of electroweak theory, for example,
the rate of baryon number violation \cite{alpha5}, and the wall velocity
of a bubble nucleated at a first order phase transition \cite{Moore_wall},
depend on the low-frequency dynamics of non-perturbative
magnetic fields.

\section {Quasiparticle scattering}

\begin {figure}
\epsfig{file=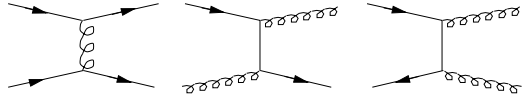,scale=0.35}
\caption
    {%
    $2\leftrightarrow 2$ particle scattering processes.
    \label {fig:1}
    }
\end {figure}

Understanding the various scattering processes which can affect
a hard quasiparticle is a prerequisite for calculating any
observable which probes quasiparticle dynamics.
These scattering processes include the basic $2 \leftrightarrow 2$
particle processes of gauge boson exchange, Compton scattering,
and pair annihilation illustrated in Fig.~\ref{fig:1}.
For a hard scattering with momentum transfer $\q \sim T$,
the explicit $g^2$ in the amplitude, plus dimensional analysis,
implies that the rate is of order $g^4 T$.
But a soft scattering, with momentum transfer $\q \sim gT$,
has a much larger $O(g^2 T)$ rate due to the $1/q^2$ behavior
of the exchanged gluon propagator, which reflects the long-range
nature of Coulomb interactions.
This enhancement of soft scattering is essentially cut off
by Debye screening at the $gT$ scale.
Note that a soft scattering event causes only a small $O(1/g)$ change
in a quasiparticle's direction, but will make an $O(1)$ change in its color.

The $O(g^4 T \ln g^{-1})$ large angle scattering rate is the same as
the two-body hard scattering rate except for the $\ln 1/g$ factor.
This arises because an $O(1)$ change in angle need not be produced
by a single hard scattering --- it may also occur via a sequence of
many soft scatterings each making a small change in the particle's direction,
which add incoherently to produce an $O(1)$ deflection.
This gives rise to a log enhancement which is cut off at the soft $gT$ scale.
And the $O[(g^2 T \ln 1/g)^{-1}]$ color coherence length is just
the inverse of the two-body soft scattering rate again up to a log
factor that arises from logarithmic sensitivity to transverse magnetic
scattering with $\q \ll gT$,
which is only cut off by non-perturbative physics \cite{gammag}.

\begin {figure}
\centerline{\epsfig{file=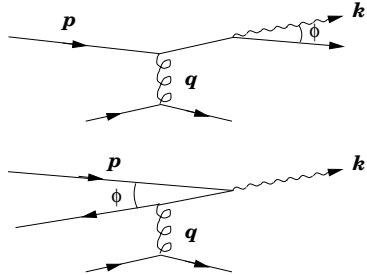,scale=0.35}}
\caption
    {%
    Near-collinear bremsstrahlung and inelastic annihilation processes.
    \label {fig:2}
    }
\end {figure}

In addition to these $2\leftrightarrow2$ particle processes,
it is also important to consider the bremsstrahlung and inelastic
annihilation processes illustrated in Fig.~\ref {fig:2}
\cite {Gelis1,Gelis2,Gelis3}.
These amplitudes contain an explicit factor of $g^3$,
so on would naively expect an $O(g^6 T)$ rate which would be
suppressed relative to the $g^4 T$ hard scattering rate.
However, if one examines the near-collinear, soft-exchange region
in which the momentum transfer $\q \sim gT$ and the angle $\phi \sim g$,
then one finds that these rates have a $1/g^4 T^4$ enhancement from
the exchanged soft gluon propagator,
a further $1/g^4 T^2$ enhancement from the nearly on-shell internal
quark propagator,
and a $g^2$ suppression from the near-collinear kinematics at the
photon vertex.
When combined with the explicit $g^6$ factors plus a $g^4 T^4$
phase space suppression from the restricted kinematics,
one finds an $O(g^4 T)$ rate --- exactly the same as the hard scattering rate.

Consequently, a hard quark (or gluon) moving through the plasma
can ``fission'' into a nearly collinear pair of hard excitations,
or ``fuse'' with a another nearly collinear hard excitation,
at a rate which is parametrically the same as the two body
scattering rate.
These processes may be regarded as $1 \leftrightarrow 2$ particle
processes which are normally forbidden by kinematics, but which
become kinematically allowed when accompanied by a soft transfer
of momentum to other particles in the system.
However, if one soft scattering with the rest of the system
can take place during a near-collinear bremsstrahlung or annihilation
process, so can two or more.
The $1/g^4 T^2$ enhancement from the internal quark propagator is a sign
that virtual intermediate states in these processes are living for a
parametrically long time of order $1/g^2 T$ --- which is not small
compared to the color coherence time.
Therefore, multiple soft scatterings occurring during these near-collinear
processes cannot be neglected and will produce an $O(1)$ suppression
in the resulting rate.
This is known as the Landau-Pomeranchuk-Migdal (LPM) effect
\cite {LP1,LP2,M1,M2}.

The net result is that a {\em leading-order} calculation of any
observable which is sensitive to the dynamics of hard quasiparticles
must correctly incorporate both the appropriate two-body scattering
processes as well as LPM-suppressed near-collinear emission processes.

\section {Photon emission rate}

\begin {figure}
\centerline{\epsfig{file=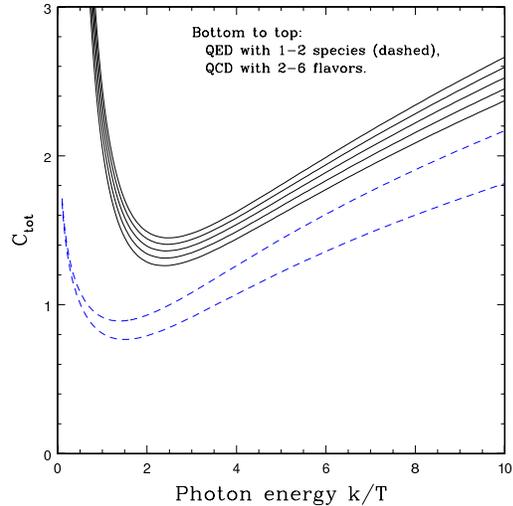,scale=0.35}}
\caption
    {%
    Total constant under the log, $C(k/T)$,
    for photon emission from hot QCD.
    (Reproduced from Ref.~\cite{AMY3}.)
    \label {fig:3}%
    }
\end {figure}

The photon emission rate from hot QCD is one quantity which has received 
considerable theoretical attention
\cite{Gelis1,Gelis2,Gelis3,Kapusta,Baier,o1,o2,o3,o4,o5,Baier_QED,Zakharov_QED},
due in part to the hope that it will
provide a useful probe in heavy ion collisions.
A complete leading-order evaluation of this emission rate
(in equilibrium) has only recently been completed \cite {AMY3}.
The differential emission rate for photons of momentum $k$ has the form
\begin {equation}
    {d\Gamma_\gamma \over dk}
    =
    {\cal B}(k) \,
    \left[ \ln (T/m_\infty) + C(k/T) + O(\gs) \right],
\end {equation}
where
$
    {\cal B}(k)
    =
    {4 \over \pi} \, \alphaEM \, \alphas \, T^2 \,
    \left( {\textstyle \sum_s} q_s^2 \right)
    k / (e^{k/T}{+}1) \,,
$
$q_s$ is the charge assignment of quark species $s$, and
$m_\infty^2 \equiv \gs^2 \, T^2/3$ is the asymptotic thermal quark mass.
The function $C(k/T)$,
shown in Fig.~\ref {fig:3},
is the ``constant under the log'';
it is a non-trivial function of $k/T$
but it is independent of the strong coupling $\gs$, whereas
$\ln (T/m_\infty) \sim \ln (1/\gs)$ since $m_\infty = O(\gs \, T)$.
The lowest order two body processes $q\bar q \to g\gamma$
and $q g \to q \gamma$ generate the $\ln (T/m_\infty)$ term
and part of $C(k/T)$;
these contributions were evaluated in \cite {Kapusta,Baier}.
The near-collinear processes only affect the
non-logarithmic term $C(k/T)$.
Their contribution to the leading-order emission rate was recently shown
to involve the solution of a non-trivial integral equation which
incorporates the effects of multiple soft scatterings during the
emission process, dynamical screening in the plasma, and thermal
corrections to the quasiparticle dispersion relations \cite {AMY2}.

\section {Transport coefficients}

Transport coefficients, such as shear viscosity or flavor diffusion constants,
are physical observables of obvious interest which are sensitive to
the dynamics of quasiparticles.
Specifically, they depend on the rates at which non-equilibrium
fluctuations in the phase space distribution
of quasiparticles relax back toward equilibrium.
And these rates depend on exactly the same scattering processes depicted in
Figs.~\ref{fig:1} and \ref {fig:2}.
Parametrically, these transport coefficients are proportional to the
large angle scattering time and hence scale as $(g^4 \ln 1/g)^{-1}$.
The fact that transport coefficients grow linearly with mean free time
is well known from the simple Drude model of conductivity;
the current induced by an applied field is proportional to the
length of time between collisions during which charges are accelerated
by the field.
It is the large angle scattering rate which is relevant,
not the much faster rate of soft scattering,
because these transport coefficients all involve the flux
of gauge invariant conserved densities
({\em i.e.}, momentum density, isospin density, etc.),
and the contribution of a quark or gluon to these fluxes is
not significantly changed if it undergoes a
scattering with tiny momentum transfer.%
\footnote
    {%
    This is not true for the ``color conductivity'' of a non-Abelian plasma,
    which is sensitive to very soft scattering processes.
    }

These transport coefficients may be formally defined by Kubo formulas
relating them to the zero frequency limit of a current-current
or stress tensor-stress tensor Wightman correlation function.
The inverse dependence of the coupling constant is an obvious sign
than an infinite number of Feynman diagrams must contribute to the
leading order result.%
\footnote
    {%
    The usual loop expansion breaks down because one is evaluating the
    correlation function at an exceptional point in momentum space,
    namely $\omega, \k \to 0$.
    }
However, the complicated set of diagrams which contribute to
the leading-order result may be summed up by a suitable integral
equation, which is precisely a linearized Boltzmann equation
(projected onto a particular symmetry channel)
for small perturbations in the quasiparticle phase space densities.

As a specific example, the flavor diffusion constant characterizing the
relaxation of fluctuations in isospin or strangeness density
(in the high temperature limit where quark masses are negligible)
has the form
\begin {equation}
    D = {\cal A} / \left\{ \gs^4 T \left[ \ln \gs^{-1} + O(1) \right] \right\},
\end {equation}
where ${\cal A}$ is a pure number
(depending on the number of active quark flavors).
Just as for the photon emission rate, the leading-log coefficient $\cal A$
is determined by $2\leftrightarrow2$ scattering processes,
while the $O(1)$ constant ``under'' the log also depends on
near-collinear emission and absorption processes in the presence
of multiple soft scatterings.
Quite a few efforts have been made to evaluate QCD transport coefficients
like $D$ in a ``leading-log'' approximation
(where one pretends that $\ln 1/\gs \gg 1$)
by solving a linearized Boltzmann equation incorporating
appropriate $2\leftrightarrow 2$ particle scattering rates 
\cite{BMPRa,heiselberg,Heiselberg_diff,BaymHeiselberg,JPT1,MooreProkopec,JPT2}.
Nevertheless, until recently almost all reported results were incorrect
due to a failure to appreciate that Compton scattering and
pair annihilation processes, in addition to gluon exchange,
contribute to the leading-log result.
This has been remedied in recent work \cite {AMY1} which found
\begin {equation}
    {\cal A}
    \simeq
    {2^4 \, 3^6 \, \zeta(3)^2 \, \pi^{-3} \over 24 + 4\nf + \pi^2} \,,
\end {equation}
for QCD with $\nf$ flavors.%
\footnote
    {%
    This is an approximate form, accurate to within a fraction of a percent,
    which results from using a one-term variational ansatz
    in the relevant integral equation \cite {AMY1}.
    }
The three terms in the denominator arise, in order,
from $t$-channel gluon exchange
between a quark and a gluon, $t$-channel gluon exchange between
two quarks, and Compton scattering or annihilation to gluons.

Analogous leading-log results for shear viscosity and electrical conductivity
may also be found in \cite {AMY1}.

Of course, a leading-log result, in which the constant under the log is
totally undetermined (and hence relative corrections suppressed
only by $1/\ln \gs^{-1}$ are neglected) is unlikely to provide
a useful prediction in any realistic theory.
At the very least, one would like to obtain a complete leading-order
result, in which neglected effects are suppressed by at least a factor of
the coupling $\gs$.
Doing so requires the correct inclusion of near-collinear gluon emission
or absorption processes analogous to the near-collinear processes
relevant for photon emission.
It should be possible to augment the linearized Boltzmann equation
with effective $1 \leftrightarrow 2$ particle scattering terms which
correctly reproduce these near-collinear reactions, and whose transition
rates incorporate the correct LPM suppression effects.
(However, this will require solving a non-trivial integral equation
just to determine the kernel to be used in another integral equation.)
Such work is currently in progress.

\section {Soft gauge field dynamics}

As noted earlier, a few important observables like the rate
of baryon number violating transitions (also known as the
``topological transition rate'') are not primarily sensitive
to the dynamics of hard quasiparticles, but instead probe
low frequency, long wavelength gauge field dynamics.
The starting point for understanding this regime is the observation
that the relevant degrees of freedom, namely modes of the gauge field
with $\k,\omega \ll T$, will have parametrically large occupation
numbers since the Bose distribution $n_b(\omega) \sim T/\omega$
for $\omega \ll T$.
Consequently, these modes may be viewed as classical fields
\cite {classical}.
But these soft modes of the gauge field are driven by
the color current generated by all the hard quasiparticles
in the plasma.
If one splits the theory into hard and soft degrees of freedom,
one may formulate a Boltzmann-Vlasov kinetic theory which describes
the propagation of ultrarelativistic hard excitations in the background
of a long wavelength classical gauge field, together with the
non-Abelian Maxwell equation $D_\nu F^{\mu\nu} = j^\mu$ characterizing
the reaction of the hard quasiparticles back on the soft gauge field.
This kinetic theory (linearized in the deviation of quasiparticle
distributions away from equilibrium) is a formulation of the well-known
hard thermal loop (HTL) effective theory~\cite {HTL1,HTL2,HTL3}.

As shown by B\"odeker \cite {Bodeker},
one may systematically integrate out the effects of fluctuations
down to a scale $\mu \ll gT$,
and derive an effective theory for the soft gauge field alone.
Of course, since hard quasiparticles propagate as nearly free particles,
with definite color, over distances up to the color coherence length
$\gamma^{-1} = O[(g^2 T \ln 1/g)^{-1}]$,
this effective theory will be non-local on the scale of $\gamma^{-1}$.
But if one restricts attention to distances large compared to the
color coherence length, or wavenumbers $\k \ll \gamma$,
then one may formulate a consistent local effective theory.
The result is a stochastic theory which simply combines Ampere's law,
Ohm's law, and the fluctuation-dissipation theorem
\cite {Bodeker,Blog1},
\begin {equation}
    {\rm D} \times {\rm B} = \sigma \, {\rm E} - \zeta \,.
\label {eq:eff}
\end {equation}
The single parameter $\sigma$ is the color conductivity,
and $\zeta$ is Gaussian noise with a variance
$\langle \zeta(x) \zeta(y) \rangle = 2 \sigma T \, \delta^4(x{-}y)$.

As noted in the introduction, the color coherence length $\gamma^{-1}$ is
parametrically {\em smaller} than the non-perturbative magnetic length
by one factor of $1/\ln g^{-1}$.
Consequently, if one applies this effective theory to 
non-perturbative $g^2 T$ scale physics, then corrections
to this effective theory will be suppressed by powers of $1/\ln g^{-1}$.
In fact, it is possible to show \cite {AY1} that corrections are suppressed by
two powers of $1/\ln g^{-1}$ provided one determines
the correct value of the color conductivity by appropriately matching
the long distance effective theory to the underlying HTL theory which
in turn is derived straight from hot QCD.
One finds (for an SU($\nc$) theory)
\begin {equation}
    \sigma^{-1}
    =
    {3 \nc \, \alpha \, T \over \mD^2}
    \left[\,
	\ln \left({\mD \over \gamma(\mu)}\right) + {\cal C}
    \,\right] ,
\end {equation}
where $\mD$ is the leading-order Debye mass (inverse screening length),
$\gamma(\mu) = \nc \, \alpha \, T \ln (\mD/\mu)$,
$\mu$ is a renormalization scale which should be chosen to be of
order $\gamma$, and ${\cal C} = 3.041$.
The leading-log coefficient was found in \cite {Bodeker,Selikov},
while the constant under the log, $\cal C$, is the result of the
very tricky matching calculation in \cite {AY2}.

In the effective theory (\ref {eq:eff}) pure dimensional analysis shows
that the the topological transition rate (per unit volume) has the form
\begin{equation}
    \Gamma = \kappa \, (\alpha T)^5 / \sigma
\end{equation}
with $\kappa$ a dimensionless pure number which depends
on the non-perturbative dynamics of the theory.
But this effective theory is a superrenormalizable, UV finite theory
which may be discretized on a spatial lattice and numerically simulated
in a completely unambiguous fashion.
(In fact, in $A_0 = 0$ gauge, the effective theory (\ref {eq:eff})
is precisely the stochastic quantization of 3$d$ Yang Mills theory.)
Such a numerical simulation was performed in \cite {Moore1},
with a result that $\kappa = 22.6 \pm 1.5$.

The topological transition rate has also been extracted from real time
simulations \cite {top-trans,moore-SEWM} of two different more microscopic
({\em i.e.}, less ``effective'') theories which are lattice versions
of the Boltzmann-Vlasov theory described earlier.
Surprising good agreement was found between the results obtained
using the long distance effective theory (\ref {eq:eff}) and both
of these more microscopic formulations.
As a result, the non-perturbative high temperature baryon violation rate
is now a satisfyingly well determined quantity \cite {moore-SEWM}.

\section {Open questions}

Many aspects of hot gauge field dynamics offer opportunities
for further progress.
Examples of problems which can be addressed using perturbative methods
include the following.
\begin {enumerate}
\item
    Perform complete leading-order evaluations of QCD transport coefficients
    such as shear viscosity, flavor diffusion, and electrical conductivity.
\item
    Evaluate bulk viscosity in hot QCD.
    Bulk viscosity does not receive a leading-log contribution
    from $2 \leftrightarrow 2$ particle processes, and at present
    there are no published results deriving even the parametric dependence
    on coupling in a gauge theory.
\item
    Compute any transport coefficient beyond leading order,
    even in a non-gauge theory.
    Can a valid effective theory still take the form of kinetic theory?
\end {enumerate}

Other natural topics involve the development or improvement
of methods for studying real time dynamics outside of perturbation theory.
To list just a few goals:
\begin {enumerate}
\item
    Test the domain of utility of leading order,
    or leading-log
    perturbative results for observables like transport coefficients.
    This has been accomplished, so far, only in one special case, namely
    the $\nf \to \infty$ limit \cite {largeNf},
    where the next-to-leading log approximation
    works surprisingly well as long as the coupling $g^2 \nf$
    is not so strong that its scale dependence becomes a dominant effect.
\item
    Develop a practical scheme for extracting transport coefficients
    from Euclidean lattice gauge theory simulations.
    This would involve extracting an estimate for the zero frequency slope
    of a spectral density from knowledge of a Euclidean correlator.
\item
    Extract more physics from real-time classical lattice gauge theory
    simulations.
    It should be possible to exhibit the presence of over-damped
    low frequency gauge field dynamics in observables
    other than the topological transition rate.
    Current efforts in this direction have found somewhat perplexing
    results \cite {AAK}.
\item
    And last but not least, develop better methods for exploring
    the real time dynamics of systems with substantial departures
    from equilibrium.
    A couple of recent steps in this direction include \cite {BMSS,Berges}.
\end {enumerate}

\newpage
\sloppy

\end{document}